\documentclass[letterpaper, 10pt, conference]{ieeeconf}

\IEEEoverridecommandlockouts
\usepackage{amsmath}
\usepackage{amsfonts}

\newtheorem{theorem}{Theorem}
\newtheorem{remark}{Remark}

\newtheorem{example}{Example}

\newtheorem{condition}{Condition}

\renewcommand{\epsilon}{\varepsilon}
\renewcommand{\theta}{\vartheta}

\newcommand*{\tran}{'}

\newcommand{\dd}[2]{\frac{\partial{#1}}{\partial{#2}}}

\newcommand{\pder}[2][]{\frac{\partial#1}{\partial#2}}
\newcommand{\R}{\mathbb{R}}
\renewcommand{\d}{\delta_x}
\DeclareMathOperator{\diag}{diag}

\begin{document}

\title{\LARGE \bf An Amendment to ``Control Contraction Metrics: Convex and Intrinsic
Criteria for Nonlinear Feedback Design''}

\author{Ian R. Manchester$^1$ and Thomas L. Chaffey$^1$%
\thanks{$^1$The authors are with the Australian Centre for Field Robotics (ACFR),
University of Sydney, NSW 2006, Australia. Emails: \texttt{ian.manchester@sydney.edu.au},\texttt{tcha4856@uni.sydney.edu.au}
.}}

\maketitle

\begin{abstract}
	We provide an amendment to the first theorem of ``Control Contraction Metrics: Convex and Intrinsic
Criteria for Nonlinear Feedback Design'' by Manchester \& Slotine in the
	form of an additional technical condition required to show
	integrability of differential control signals.  This technical condition is shown to be satisfied under the original assumptions if the input matrix is constant rank, and also if the strong conditions for a CCM hold. However a simple counterexample shows that if the input matrix drops rank, then the weaker conditions of the original theorem may not imply stabilizability of all trajectories. The remaining claims and illustrative examples of
	the paper are shown to remain valid with the new condition. 
\end{abstract}

\section{Introduction}

This note considers dynamical systems of the form
\begin{IEEEeqnarray}{rCl}
	\dot{x} &=& f(x, t) + B(x, t)\,u \label{dynamics}
\end{IEEEeqnarray}
where $x\in\mathbb{R}^n$, the drift $f$ and input matrix $B$ are smooth functions and $u \in \mathbb{R}^m$.
The differential dynamics defined along trajectories of \eqref{dynamics} are given by
\begin{IEEEeqnarray}{rCl}
	\dot{\delta_x} &=& A(x, u, t)\,\delta_x + B(x, t)\,\delta_u,
	\label{differential}
\end{IEEEeqnarray}
where $A = \dd{f}{x} + \sum_{i=1}^n \dd{b_i}{x} u_i$, $b_i$ represents the
$i^\text{th}$
column of $B$ and $u_i$ represents the $i^\text{th}$ element of $u$.

The first theorem of the paper \cite{Manchester2017} was as follows:
\begin{theorem}\label{TH1}
	Suppose there exists a metric $M(x,t)$ such that there exist $\alpha_1,
	\alpha_2 \geq 0$ with $\alpha_1 I \leq M(x,t) \leq \alpha_2 I$. Suppose that for every $t$, $x$, $u$ and $\delta_x \neq 0$.  
	\begin{IEEEeqnarray*}{rCl}
		\delta_x\tran MB = 0 \implies \delta_x\tran\left( \dot{M} + A\tran M
		+ MA + 2\lambda M\right)\delta_x < 0\\
		\IEEEyesnumber \label{th1}
	\end{IEEEeqnarray*}
	for some $\lambda > 0$.	
%Furthermore, suppose that for
%	all compact subsets $X \subset \mathbb{R}^n$, for all $x\in X$, and for all
%	compact subsects $Y \subset \mathbb{R}^n$ not containing 0, for all $\delta_x
%	\in Y$, the ratio
%	\begin{IEEEeqnarray}{C}
%		\frac{\dd{\dot{V}}{u}}{\dd{V}{\delta_x} B B\tran
%		\dd{V}{\delta_x}\tran} \label{ratio}
%	\end{IEEEeqnarray}
%	is bounded, where
%	\begin{IEEEeqnarray*}{rCl}
%		\dot{V} &=& \dd{V}{t} + \dd{V}{x}\left(f + Bu\right) +
%		\dd{V}{\delta_x}A\delta_x.
%	\end{IEEEeqnarray*}
	Then the system \eqref{dynamics} is
	\begin{itemize}
		\item universally exponentially open-loop controllable;
		\item universally exponentially stabilisable via sampled data
			feedback with arbitrary sample times;
		\item universally exponentially stabilisable via continuous feedback
			defined almost everywhere, and everywhere in a neighbourhood
			of the target trajectory;
	\end{itemize}
	all with rate $\lambda$ and overshoot $R = \sqrt{\alpha_2/\alpha_1}$.
\end{theorem}
For definition of these terms see \cite{Manchester2017}.

%The condition that (\ref{ratio}) is bounded is true for any system meeting the
%conditions of \cite[Prop. 1]{Manchester2017} or the strong
%conditions given in \cite[Sec. III. A.]{Manchester2017}.  

Defining $V(x,\delta,t) =
	\delta_x\tran M(x,t) \delta_x$, we can state an equivalent formulation of \eqref{th1} in terms
of $V$, which can be thought of as a differential version of the Artstein-Sontag condition for a control Lyapunov function \cite{Sontag1989}:
	\begin{IEEEeqnarray*}{rCl}
		\dd{V}{\delta_x}B &=& 0 \implies \dd{V}{t} + \dd{V}{x}\left(f + Bu\right) +
		\dd{V}{\delta_x}A\delta_x < -2\lambda V.
	\end{IEEEeqnarray*}
	
However, the following simple counterexample shows that Theorem \ref{TH1} as stated in \cite{Manchester2017} is not correct:	
\begin{example}\label{counterexample}
	Let $\mathcal{M} = \mathbb{R}$ and consider the system \eqref{dynamics} with $f(x)=-x, B(x)=x^2$, i.e.
	\begin{IEEEeqnarray*}{rCl}
		\dot{x} = -x + x^2 u,
	\end{IEEEeqnarray*}
	and consider the candidate metric $V = \delta_x^2$. We have
	$
		\dd{V}{\delta_x} B = 2\delta_x x^2$ and $\dd{V}{\delta_x} \left( \dd{f}{x} + \dd{B}{x}u\right)\delta_x =
		-2\delta_x^2 + 4\delta_x^2 u x.$
	Hence the system and Lyapunov function meet the requirement~(\ref{th1}).
	%However, computing the ratio~(\ref{ratio}) gives $1/x^3$, which is unbounded
	%as $x \rightarrow 0$.  This means the control signal cannot be integrated.
However, if the system is at $x=0$, then note that the control input has no effect. And $x(t)=1\,\forall t, u(t)=1\,\forall t$ is a valid solution which can not be reached from $x=0$. Hence the system is not universally open-loop controllable.
\end{example}
\section{An Amended Theorem}
For conciseness we make the following definitions:
\begin{align}
  H_i(x,t) &:= \partial_{b_i}M +\pder[b_i]{x}'M+M\pder[b_i]{x}\\
Q(x,t) &:= M(x,t)B(x,t)B(x,t)'M(x,t)
\end{align}
Note that $\d'H_i(x,t)\d= \pder[\dot V]{u_i}$, and $Q(x,t)$ is positive semidefinite by construction.
Consider the following additional condition:
\begin{condition}\label{newcond}
	There exists a continuous function $\psi(x,\d, t):\R^n\times\R^n\times[0,\infty)\to[0,\infty)$ such that for each $i = 1, 2, ... m$
\begin{equation}\label{eqn:psi}
  |\d'H_i(x,t)\d|\le \psi(x,\d,t)\d'Q(x,t)\d 
\end{equation}
for all $x,\d,t$.\hfill $\Box$
\end{condition}
In terms of $V$, \eqref{eqn:psi} can be equivalently written as
\begin{equation}\label{eqn:V_psi}
\left|\frac{d \dot V}{du_i}\right|\le \psi(x,\d,t)\pder[V]{\d}BB'\pder[V]{\d}'.
\end{equation}

The main result of this note is the following:
\begin{theorem}\label{thm:corrected}
	Suppose that the conditions of Theorem \ref{TH1} are satisfied, and in addition Condition \ref{newcond} is satisfied, then the claimed implications of Theorem \ref{TH1} are true.\hfill $\Box$
\end{theorem}

The proofs of this and all following results are in the appendix.

\begin{remark}
Condition \eqref{th1} already implies that
\begin{equation}\label{eqn:CCM_u}
\d'Q(x,t)\d=0\Longrightarrow \d'H_i(x,t)\d=0,
\end{equation}
since the left-hand-side in \eqref{th1} is true precisely when $\d'Q(x,t)\d=0$, and the right-hand-side of  \eqref{th1} is affine in $u$, and hence only sign-definite if the terms multiplying $u_i$ are zero, which are precisely  $ \d'H_i(x,t)\d$. The stronger condition \eqref{eqn:psi} implies that $\d'Q(x,t)\d$ des not go to zero more rapidly than $\d'H_i(x,t)\d$.
\end{remark}

\begin{remark}
  In Example \ref{counterexample}, ${\d'H_i\d}=4\d^2x$ and ${\d'Q_i\d}=4\d^2x^4$, so we would need
$x\le\psi(x)x^4$, which would clearly require $\psi$ to blow up at zero, contradicting $\psi$ being well-defined and continuous on $\R$. 

\end{remark}

%We now provide a proof that the ratio \eqref{eqn:psi} guarantees
%integrability of differential control signals.  Every other aspect of the proof of
%Theorem~\ref{TH1} remains unchanged and we refer the reader to
%\cite[Appendix]{Manchester2017} for details.

\section{Sufficient Conditions and Implications}

In this section we first note note two important special cases in which Condition \ref{newcond} is automatically satisfied. 

\begin{theorem}\label{thm:strong}
	Suppose the Strong Conditions C1 and C2 of \cite[Sec. III.A]{Manchester2017} hold. Then Condition \ref{newcond} holds.
\end{theorem}

We note that all illustrative examples in \cite{Manchester2017} used the strong conditions.

\begin{theorem}\label{thm:rank}
	Suppose a system satisfies the conditions of Theorem \ref{TH1} and, in addition, the matrix $B(x,t)$ has constant rank $m$ for all $x,t$, then Condition \ref{newcond} holds. \hfill $\Box$
\end{theorem}

\begin{remark}
  In Example \ref{counterexample}, $\tilde D(x)=x^4$ which is not strictly positive for all $x$. \hfill $\Box$
\end{remark}

We conclude this note by confirming the validity of results of Section III.B, IV.B, and IV.C of \cite{Manchester2017} with the new condition \ref{newcond}.

\begin{theorem}\label{thm:convex}
Condition \eqref{newcond} is convex under the same change of variables used in \cite[Sec. III.B]{Manchester2017}. \hfill $\Box$
\end{theorem}

\begin{theorem}\label{thm:intrinsic} Condition \eqref{newcond} is invariant under the   smooth coordinate changes and affine feedback transformations considered in \cite[Theorem 2]{Manchester2017}. \hfill $\Box$
\end{theorem}

We also note that \cite[Corollary 1]{Manchester2017} on necessity for feedback linearizable systems follows as in \cite{Manchester2017} from Theorems \ref{thm:strong} and \ref{thm:intrinsic}, and the fact that stabilizable linear systems satisfy the strong conditions C1 and C2.

%Hence we have
%\begin{IEEEeqnarray*}{rCl}
%	\frac{\dd{\dot{v}}{u}}{\dd{V}{\delta_x} B \beta \beta\tran B\tran
%	\dd{V}{\delta_x}\tran} &\leq& \frac{1}{\gamma}\frac{\dd{\dot{v}}{u}}{\dd{V}{\delta_x} B  B\tran
%	\dd{V}{\delta_x}\tran},
%\end{IEEEeqnarray*}
%which is bounded.
\appendix

{\em Proof of Theorem \ref{thm:corrected}}
Following \cite{Manchester2017}, define
\begin{IEEEeqnarray*}{rCl}
	a(x, \delta_x, u, t) &=& \dd{V}{t} + \dd{V}{x}\,(f + Bu) + \dd{V}{\delta_x}\,A\delta_x +
	\alpha(V)\\
	b(x, \delta_x, t) &=& \dd{V}{\delta_x}\,B\,B\tran\dd{V}{\delta_x}\tran.\\
	\rho(x, \delta_x, u, t) &=& \left\{ \begin{array}{c l}
						0 & \text{if}\; a < 0\\
						\frac{a + \sqrt{a^2 + b^2}}{b} &
						\text{otherwise.}
	\end{array}\right.
\end{IEEEeqnarray*}
%The differential feedback control is given by
%\begin{IEEEeqnarray}{rCl}
%	k_\delta(x, \delta_x, u, t) &=& -\rho(x, \delta_x, u, t)\,B(x, t)\tran\dd{V(x,
%	\delta_x, t)}{\delta_x}\tran.\label{kd}
%\end{IEEEeqnarray}
%	We now show that the differential control signal (\ref{kd}) is integrable
%	along regular curves in $\mathbb{R}^n$.  That is, for any regular curve
%	$c:[0,1] \rightarrow \mathbb{R}^n$ and any $u_0\in \mathbb{R}^m$,
%	$t\in\mathbb{R}^+$, a unique solution of the following integral equation
%	exists on $s\in [0,1]$:
%	\begin{IEEEeqnarray}{rCl}
%		v(s) &=& u_0 + \int^s_0 k_\delta\left(c(s), c_s(s), v(s),
%		t\right)\mathrm{d}s.\label{int}
%	\end{IEEEeqnarray}
%The condition \eqref{th1} implies either $b > 0$ or $a < 0$.  It follows from 
%	\cite[Th. 1]{Sontag1989} that $\rho$ is smooth for all $x$, $u$ and $\delta_x
%	\neq 0$.
%	%, and the apparent discontinuity at $b = 0$ is removed by setting
%	%$\rho = 0$ when $b = 0$
%  Smoothness of $\rho$ implies smoothness of
%	$k_\delta$ when $\delta_x \neq 0$, which is the case in Equation~(\ref{int})
%	where $\delta_x$ is set to $c_s(s)$, which is non-zero by regularity of $c(s)$.
%
%	It follows from \cite[Th. 3.2]{Khalil2002} that a unique solution to (\ref{int})
%	exists if $k_\delta$ is a globally Lipschitz function with respect to its
%	third argument for $s\in[0, 1]$.  As $B$ and $V$ are continuously
%	differentiable and smooth respectively, the product $B(\partial V/\partial
%	\delta_x)$ is bounded on
%	closed intervals. 
	
As in \cite{Manchester2017}, if $\rho$ is globally Lipschitz in $u$ on any bounded subset of $x, \delta_x$ for fixed $t$, then the differential control law is integrable and result follows. As $\rho$ is smooth, it is globally
	Lipschitz if $|\dd{\rho}{u}|$ is bounded.  
	This is clear for $b \leq 0$.  For $b > 0$, noting that the only dependence
	$\rho$ has on $u$ is via $a$, we have
	\begin{IEEEeqnarray*}{rCl}
		\dd{\rho}{u}		&=& \frac{1}{b}\dd{a}{u} \left(1 + \frac{a(u)}{\sqrt{a^2(u) +
		b^2}}\right).
	\end{IEEEeqnarray*}
	Since $a = 0 \implies b > 0$ and $a$ is affine in $u$, the only term that can
	be unbounded is $(1/b)(\partial a/\partial u)$.  However, by construction
	$\partial a/\partial u_i= \d'H_i\d$ and $b=\d'Q\d$. Hence, by Condition
	\ref{newcond}, $|(1/b)(\partial a/\partial u_i)|\le \psi(x,\delta_x, t)$ and
	hence $\rho$ is globally Lipschitz in $u$. The result follows. \hfill $\Box$
	
	{\em Proof of Theorem \ref{thm:strong}:}
	Condition C2  is precisely that $\d'H_i\d=0$ for all $i$, hence \eqref{eqn:psi} is true with $\psi =0$. \hfill $\Box$

{\em Proof of Theorem \ref{thm:rank}:}
In this case, clearly $Q(x)$ has rank $m$ for all $x$, since $M(x)>0\, \forall x$. We will show that in this case, \eqref{eqn:CCM_u} implies existence of $\psi(x)$ satsifying \eqref{eqn:psi}.

Since $Q$ is symmetric positive-semidefinite, at each $x$ the set $\{\d:\d' Q(x)\d=0\}$ is a subspace spanned by eigenvectors of $Q(x)$ with zero eigenvalues. I.e. there is a decomposition
$
Q (x)= V(x) D(x) V(x)'
$
with $D = \diag(d_1, ..., d_m, 0, ..., 0)$, $d_i>0$ for $i=1,..,m$, $V'V=I$, $V = [\tilde V, \, V_0]$, $\tilde V\in\R^{n\times m}, V_0\in\R^{n\times (n-m)}$. %I.e. columns of $\tilde V$ are the eigenvectors with non-zero eigenvalues, and columns of  $V_0$ are those with zero eigenvalues. 
Keeping only non-zero terms in the expansion, we can write equivalently
$
Q (x)= \tilde V(x) \tilde D(x) \tilde V(x)'
$
where $\tilde D = \diag(d_1, ..., d_m)$.

The assumption \eqref{eqn:CCM_u} states that for  $\d$ in the span of $V_0$, $\d' H_i(x)\d$ is also zero, hence columns of $V_0(x)$ are also eigenvectors of $H_i(x)$ with eigenvalue 0. And since $H_i$ is symmetric, by orthogonality of eigenvectors, the non-zero eigenvalues of $H_i(x)$ (if any) have eigenvectors in the span of the columns of $\tilde V(x)$. That is, we can write
$
H_i(x) = \tilde V(x) \tilde H_i(x) \tilde V(x)'
$
for some symmetric $\tilde H_i(x)$. 

Now, the required bound \eqref{eqn:psi} can be rewritten as
$
|\d'\tilde V \tilde H_i \tilde V'\d|\le \psi(x)\d'\tilde V \tilde D \tilde V'\d$ which holds with
$
\psi(x) = 
\frac{\sigma_{\max}( \tilde H_i(x))}{d_m(x)}
$
where $\sigma_{\max}$ denotes  the maximum singular value. Since $d_m$ is strictly positive for all $x$, this ratio is always finite. Furthermore, $\psi$ is continuous by continuity of eigenvalues with respect to matrix elements and positivity of $d_m$. \hfill $\Box$

{\em Proof of Theorem \ref{thm:convex}}
Under the change of coordinates $W(x,t) = M^{-1}(x,t)$, $\eta = M(x, t)\delta_x$, \eqref{eqn:psi} becomes
\begin{IEEEeqnarray*}{rCl}
	\left|\eta\tran\left(W \dd{b_i}{x}\tran + \dd{b_i}{x} W - \partial_{b_i}
	W\right)\eta\right|\le\psi(x,\d,t)\eta\tran B B\tran \eta.
\end{IEEEeqnarray*}
which is jointly convex in $W, \psi$. \hfill $\Box$

{\em Proof of Theorem \ref{thm:intrinsic}:}
Invariance under changes of coordinates for the state is clear from the coordinate-free representation \eqref{eqn:V_psi}. We now show it is invariant under affine control laws,
that is, $u = \alpha(x) + \beta(x) v$, where $\beta$ is a smooth, nonsingular
$m\times m$ matrix function. I.e., assuming \eqref{eqn:V_psi} we have for the original representation:
\[
\left|\tfrac{\partial \dot V}{\partial u}\right|_\infty \le  \psi(x,\d,t)\tfrac{\partial V}{\partial \d}BB'\tfrac{\partial V}{\partial \d}'
\] 
and we wish to prove, for the transformed system, that
\begin{equation}\label{eq:trans_bound}
\left|\tfrac{\partial \dot V}{\partial v}\right|_\infty= \left|\tfrac{\partial \dot V}{\partial u}\beta\right|_\infty \le \bar\psi(x,\d,t)\tfrac{\partial V}{\partial \d}B\beta\beta'B'\tfrac{\partial V}{\partial \d}',
\end{equation}
for some $\bar\psi$. Now $|\pder[\dot V]{u}\beta|_\infty\le \kappa|\pder[\dot V]{u}\beta|_\infty$, where $\kappa$ is the maximum $\ell^1$ norm of a row of $\beta$, and $
	\gamma\dd{V}{\delta_x} B B\tran  \dd{V}{\delta_x}\tran \le \dd{V}{\delta_x} B \beta \beta\tran B\tran \dd{V}{\delta_x}\tran$
where $\gamma(x)$ is the smallest eigenvalue of $\beta \beta\tran$, and $\gamma(x)>0$ for all $x$ since $\beta$ is non-singular everywhere. Hence \eqref{eq:trans_bound} is true with $\bar\psi = \frac{\kappa}{\gamma}\psi$.
\hfill $\Box$
\bibliographystyle{IEEEtran}
\bibliography{paper}{}

\begin{thebibliography}{1}
\providecommand{\url}[1]{#1}
\csname url@rmstyle\endcsname
\providecommand{\newblock}{\relax}
\providecommand{\bibinfo}[2]{#2}
\providecommand\BIBentrySTDinterwordspacing{\spaceskip=0pt\relax}
\providecommand\BIBentryALTinterwordstretchfactor{4}
\providecommand\BIBentryALTinterwordspacing{\spaceskip=\fontdimen2\font plus
\BIBentryALTinterwordstretchfactor\fontdimen3\font minus
  \fontdimen4\font\relax}
\providecommand\BIBforeignlanguage[2]{{%
\expandafter\ifx\csname l@#1\endcsname\relax
\typeout{** WARNING: IEEEtran.bst: No hyphenation pattern has been}%
\typeout{** loaded for the language `#1'. Using the pattern for}%
\typeout{** the default language instead.}%
\else
\language=\csname l@#1\endcsname
\fi
#2}}

\bibitem{Manchester2017}
I.~R. Manchester and J.-J.~E. Slotine, ``{Control Contraction Metrics: Convex
  and Intrinsic Criteria for Nonlinear Feedback Design},'' \emph{IEEE
  Transactions on Automatic Control}, vol.~62, no.~6, pp. 3046--3053, 2017.

\bibitem{Sontag1989}
E.~D. Sontag, ``{A 'universal' construction of Artstein's theorem on nonlinear
  stabilization},'' \emph{Systems and Control Letters}, vol.~13, no.~2, pp.
  117--123, aug 1989.

\end{thebibliography}

\end{document}